# Who Is in Mind Matters: Attachment Representations in Early Childhood Synchronize Child-Adult Interacting Brains

Ruxin Su[1,2,3#], Jiayang Xu[4#], Saishuang Wu[5,6], Haiwa Wang[6], Yamin Li[4,7], Zihan Yang[8], Yuqi Liu[9], Jieqiong Liu[6], Shanbao Tong[4], Yunting Zhang[10]*, Xiaoli Guo[4]*, Fan Jiang[6,11]*

1 Shanghai Key Laboratory of Child Brain and Development, National Children's Medical Center, Shanghai Children's Medical Center, School of Medicine, Shanghai Jiao Tong University, Shanghai, China

2 School of Psychology, Shanghai Jiao Tong University, Shanghai, China

3 Center of Yuanshen Rehabilitation Institute, Shanghai Jiao Tong University School of Medicine, Shanghai, China

4 School of Biomedical Engineering, Shanghai Jiao Tong University, Shanghai, China

5 Department of Child Health Care, Children's Hospital, Zhejiang University School of Medicine, National Clinical Research Center for Child Health, Hangzhou, China

6 Department of Developmental and Behavioral Pediatrics and Pediatric Translational Medicine Institute and Shanghai Key Laboratory of Child Brain and Development, National Children's Medical Center, Shanghai Children's Medical Center, Shanghai Jiao Tong University School of Medicine, Shanghai, China

7 Department of Computer Science, Vanderbilt University, Nashville, TN, USA

8 Beijing No. 171 High School, Beijing, China

9 Global Institute of Future technology, Shanghai Jiao Tong university

10 Child Health Advocacy Institute, National Children's Medical Center, Shanghai Children's Medical Center, Affiliated to School of Medicine Shanghai Jiao Tong University, Shanghai, China

11 Shanghai Center for Brain Science and Brain-Inspired Technology, Shanghai, China

**# Ruxin Su and Jiayang Xu contributed equally to this work.**

*** Corresponding authors:**

Dr. Yunting Zhang at Child Health Advocacy Institute, National Children's Medical Center,

Shanghai Children's Medical Center, affiliated to School of Medicine Shanghai Jiao Tong

University, Shanghai, China, e-mail: edwinazhang@shsmu.edu.cn.

Dr. Xiaoli Guo at School of Biomedical Engineering, Shanghai Jiao Tong University, Shanghai, China, e-mail: meagle@sjtu.edu.cn.

Dr Fan Jiang at National Children's Medical Center, Shanghai Children's Medical Center, Shanghai Jiao Tong University School of Medicine, Shanghai, China, e-mail: fanjiang@shsmu.edu.cn.

**Fundings**

This work was supported by the Fundamental Research Funds for the Central Universities (No. YG2023QNB21, No. YG2025LC01), the National Natural Science Foundation of China (No. 62371285, No. U23A20170, No. 24dz2260100), the Innovative Research Team of High-level Local Universities in Shanghai (No. SHSMU-ZDCX20211900), Project of Yuanshen Rehabilitation Institute, Shanghai Jiao Tong University School of Medicine (Grant No. yskf1-23-1107-1[to QF])

**Acknowledgements**

The authors thank Jieyuan Zhang for assistance with data analysis, and Xiaoyan Zhang, Weiwei Cai, Yijie Hong, Jiawen Zhang, Yufeng Liu, Xiner Tan, Xintong Chen, Siyao Cao, Ruili Ma, and Mengfei Zhou for their support in data collection. The authors also sincerely thank Dr. Muming Pu and Dr. Liping Wang for their valuable assistance with manuscript writing.

## Abstract

Human attachment is distinguished by enduring internalized representations that shapes neurodevelopment and social-emotional functioning. However, as unobservable inner processes mixed with social cues and partner-specific factors, the neurocognitive mechanisms of these representations during real-time interaction remain unclear. Using a novel Remote Partner-Belief Manipulation paradigm in 40 child-mother-stranger trios, we experimentally isolated attachment representations in 3-4-year-olds by manipulating children's partner-belief during remote cooperation. The inner processes were captured from synchrony between partners' EEG, showing that children's mother-partner belief, regardless of the actual partner, significantly enhanced interbrain synchrony. This partner-belief modulation concentrated on children's P4 channel (overlaying the attachment-designated right temporoparietal junction), where synchrony strength correlated to attachment security and children's response acceleration due to mother-partner belief. These findings established attachment representations as an independent, endogenous driver of interbrain synchrony, potentially via children's heightened attention towards their attachment figure, implying the role of symbolic attachment activation when separation.

# Introduction

Attachment, a defining feature of mammalian social life shaped through early caregiver-offspring interactions, is rooted in conserved dopamine-oxytocin interactions within neurobiological systems[1,2]. During human evolution, the integration of subcortical with cortical networks gradually enabled the emergence of a representation-based attachment system, making attachment increasingly selective and enduring[1]. Through repeated caregiver-child interactions during early sensitive periods, children progressively internalize repeated interaction patterns into stable mental representations — internal working model[3]. Once established, these endogenous representations are theoretically regarded as the core and enduring component of attachment, operating autonomously beyond the immediate behavioral cues that initially shaped them, thereby outliving mother-present contexts and scaffolding durable, diverse close relationships[4–6]. However, despite their central role in attachment theory, these representations remain theoretically elusive constructs, abstract and empirically intangible, posing persistent challenges to capture the real-time inner process directly.

With advances in hyperscanning techniques, recent studies have indicated that interbrain synchrony can serve as an index of interpersonal psychological processes[7,8], by virtue of its direct measurement of dyadic neural activity. Specifically in the context of mother-child relationships, interbrain synchrony has been suggested as a neural correlate of attachment, which can be evidenced by consistently higher levels of neural synchrony in mother-child dyads compared to stranger-child dyads across various settings, including naturalistic viewing[9], free play[10,11], and cooperative tasks[12]. These findings underscored the unique text of the mother-child relationship, implying the influence of the hierarchical nature of social relationships and the facilitating role of psychological bonding in mother-child interaction.

However, almost all existing studies had children interacting with their mothers or strangers in shared physical environments, and the communicative cues they received differed across different partners, introducing significant confounding effects on neural synchrony measurements. Notably, even subtle differences in social signals, such as chemosignals[10], emotion valence[13], affectionate touch[14], gaze signals[15], as well as joint attention[16], can significantly modulate interbrain synchrony. Coupled with superior maternal caregiving profiles[17], such cues are typically more passionate and frequent in mother-child interactions[10,18], potentially contributing to higher levels of interbrain synchrony. Although some studies have attempted to control communicative cues — for example, by requiring partners to keep silent[12] or covering participants' hands during keypress responses[19] — interactions in shared physical spaces cannot fully eliminate all social cues like maternal chemosignals, proximity and emotional valence. Thus, it remains challenging to distinguish whether heightened interbrain synchrony reflects stable endogenous attachment representations or merely transient neural entrainment to context-dependent social cues.

Moreover, beyond overt communicative signals, interbrain synchrony may also be shaped by other endogenous factors independent of the children's current psychological state. These encompass genetic relatedness[20,21], shared knowledge[22], and common memories[23], along with mothers' heightened attention and attachment-related neurobiological responses toward their own children[24,25]. To isolate the specific contribution of attachment representations per se to interbrain synchrony, it is essential to adopt novel paradigms to experimentally manipulate mental representations of attachment while controlling for alternative synchrony sources.

Therefore, with the aim of validating the hypothesis that child-to-mother attachment representations would causally enhance interbrain synchrony, we developed a Remote Partner-Belief Manipulation paradigm, based on a seven-piece tangram cooperation task. In this paradigm, the influence of social cues was minimized by physically isolating children from partners (mother/stranger) using a remote setting, and the link between child-to-mother attachment representations and their interbrain synchrony was investigated by manipulating the children's believed partner through veridical or deceptive notifications, i.e., children were either correctly informed about their partner or misled into believing that their partner was the deceptive one. This design of manipulation was inspired by the 'creating an illustration of interaction partner' method adopted by Caruana et al. to explore how human agency beliefs influence behavior during virtual social interactions[26]. Through such manipulation, we established a 2 × 2 design with two orthogonal factors: (1) children's *believed partner* (*believed mother* vs. *believed stranger*) and (2) *actual partner* (*actual mother* vs. *actual stranger*) (Fig. 1). We recruited 40 children aged 3-4 years along with their mothers and 40 unfamiliar female adults, constituting 40 child-mother-stranger trios. Their EEG signals during the experiment were recorded simultaneously to capture the interbrain synchrony between children and their actual partner. Notably, ages 3–4 coincide with the preparatory period for preschool entry, during which children's primary social bonds remain centered on the main caregiver, while stable attachment representations have already been established. Considering the confounding effects of role switching on neural synchrony[12,27], a previously employed side-by-side cooperation paradigm was used as an initial step to validate the role-dependent neural coordination in EEG frequency.

Based on this paradigm, the causal relationship between child-to-mother attachment representations and interbrain synchrony could be established if the following pieces of evidence are derived: (1) Mother-child dyads would exhibit higher interbrain synchrony than stranger-child dyads even when partner-specific social cues are minimized by the remote design; (2) Interbrain synchrony would be enhanced when children believe they are interacting with their mother versus a stranger, even though the actual partner is held constant across conditions; (3) The interbrain connections modulated by partner belief would be within children's attachment-related brain regions, such as the prefrontal and temporal-parietal regions[10,12,28]; and (4) The magnitude of partner-belief modulation of interbrain synchrony would scale with attachment-related relationship scales. Beyond neural evidence, we also compared the task performance across conditions to further explore the effects of internal attachment representations at the behavioral levels.

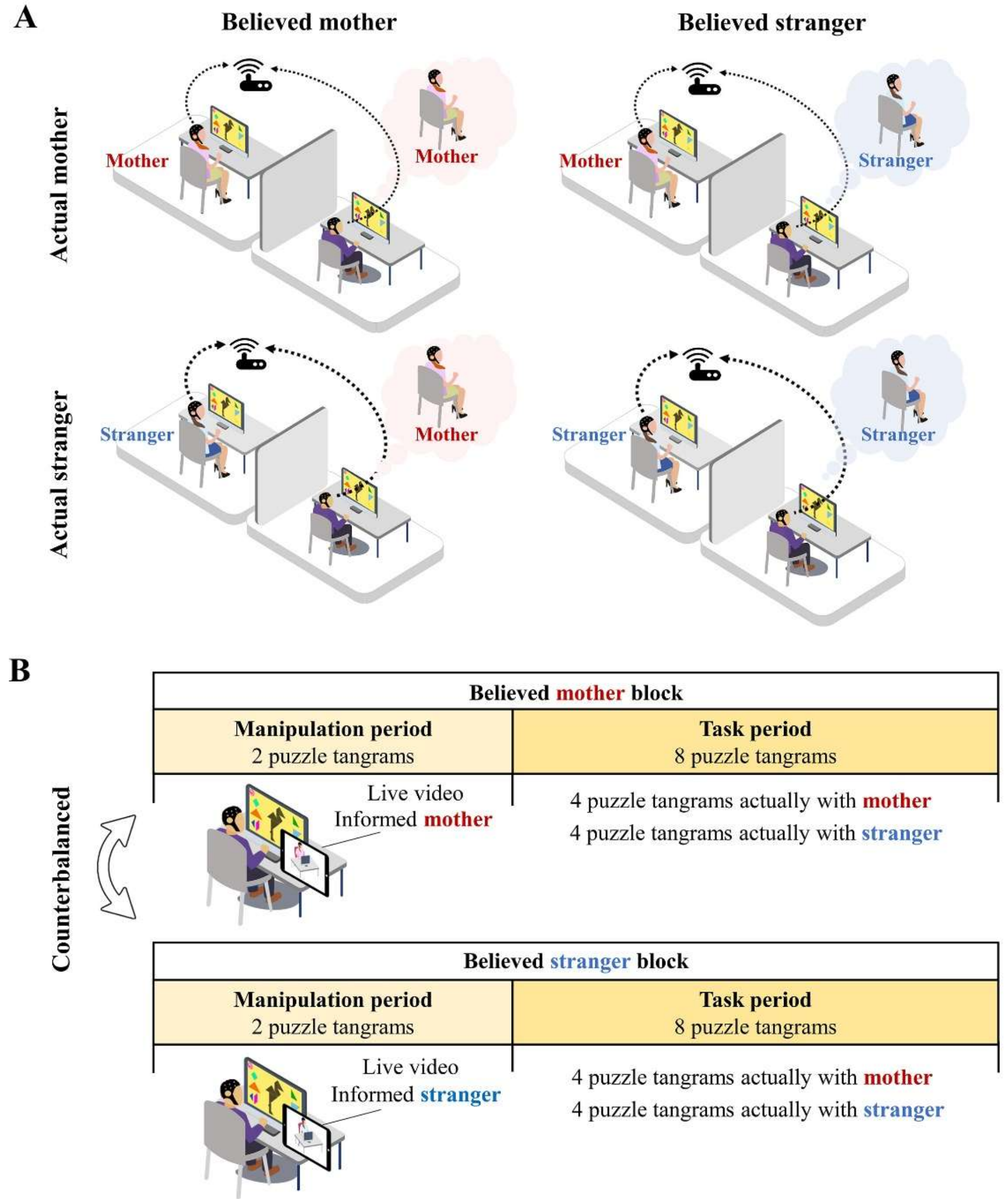

**Fig. 1 Experimental protocol**. (A) The experiment implemented a 2×2 factorial design crossing *believed partner* with *actual partner*. A child, along with his/her mother or a stranger, was seated in two adjacent, sound-proofed rooms and cooperated remotely to complete the seven-piece tangram puzzle-solving video tasks. Each child completed puzzle tasks under four cooperative conditions: (1) *believed mother - actual mother*: cooperated with mother and truly informed of the partner's identity; (2) *believed stranger - actual mother*: cooperated with mother but misidentified partner as a stranger; (3) *believed mother - actual stranger*: cooperated with stranger but misidentified partner as mother; and (4) *believed stranger – actual stranger*: cooperated with stranger and truly informed of the partner's identity. (B) The experimental paradigm. The experiment included 2 blocks (1 block per believed partner). At the beginning of each block (manipulation period), the child was instructed to cooperate with the informed partner to complete 2 puzzle tangrams for practice, during which a live video of the informed partner was shown to the child. Then after practice, the live video was removed, and the main task began, with half (4 tangrams) actually cooperating with mother and the other half (4 tangrams) with stranger. The actual partner (mother or stranger) for each tangram was randomly assigned.

# Results

## Behavioral results

Four children failed to complete the entire experimental procedure. For the remaining 36 children, we calculated their reaction time (defined as the time interval between the completion of partner's piece placement and the initiation of the children's next piece movement) and movement duration (defined as the time elapsed from the initiation to the successful placement of a geometric piece) during the puzzle tasks (Fig. 2A). These behavioral metrics were then compared across different conditions using 2 × 2 repeated-measures ANOVAs. The main effect of *believed partner* was significant on children's reaction time (F (1,35) = 6.383, $p$ = .016, $\eta_p^2$ = .154, Fig. 2B), exhibiting a modulation on children's attention through partner-belief manipulation. Specifically, when children believed to be interacting with their mothers, their reaction times were significantly reduced, whether the actual partner was mother or partner (*believed mother - actual mother* = 1.25 ± 0.052 s, *believed stranger - actual mother* = 1.48 ± 0.104 s, *believed mother - actual stranger* = 1.33 ± 0.060 s, *believed stranger - actual stranger* = 1.46 ± 0.095 s, Fig. 2B). However, neither *believed partner, actual partner*, nor their interaction showed statistically significant effects on the children's movement duration (all $p$ > .312, $\eta_p^2$ < .029, Fig. 2C).

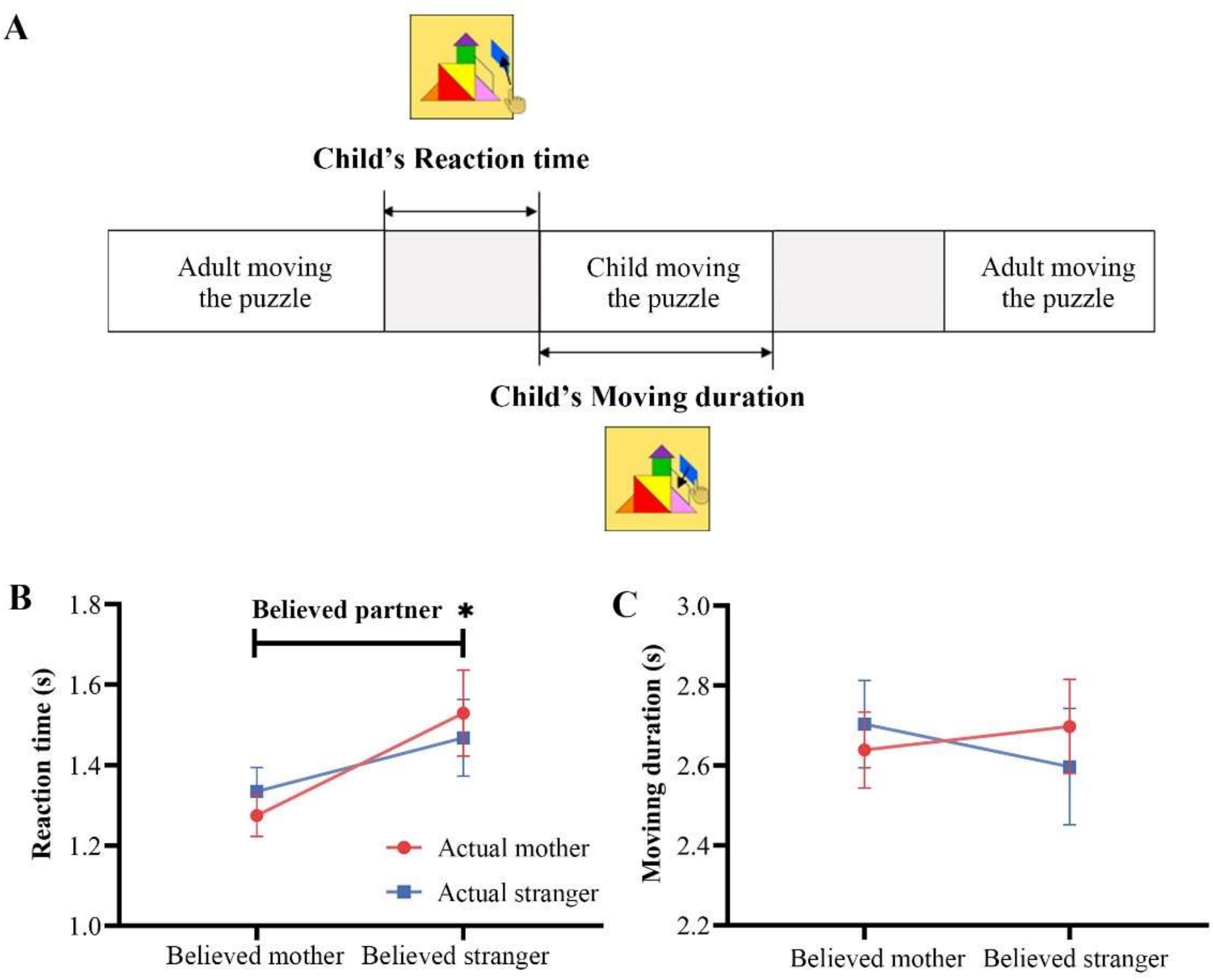

**Fig. 2 Children's behavioral performance across different conditions.** (A) An illustration of how to calculate children's moving duration and reaction time. (B) Children's moving duration, and (C) reaction time across different conditions. $^{*}p$ <.05.

**Global interbrain network metrics**

Three child-mother-stranger trios were further excluded in the interbrain analyses due to the poor EEG quality. For the included 33 trios, we constructed the interbrain networks between children and their actual partners within both theta and alpha bands using the Phase Locking Value (PLV)[29]. These bands have been consistently shown to be responsible for the adult-child interbrain communication[30,31]. Noticing that children aged 3-4 years hardly kept motionless during the observation phase (see Method and Fig. S1), we only used the EEG data when adults observed children moving the puzzles for further analysis, to exclude the potential contamination of excessive movements on interbrain synchrony. The side-by-side experiment confirmed that during this role context, the mother-child interbrain network was more densely (quantified as the density of interbrain network in group level, IBD) and strongly (quantified as the mean strength of interbrain connections in individual level, IBS) connected than that between stranger and child specifically in the theta band (Fig. S2). These findings well replicated our prior results from another sample[12].

Similar to the side-by-side experiment, remote cooperation also showed enhanced theta-band interbrain synchrony in mother-child dyads (IBD = 0.056, IBS = 0.427 ± 0.005) compared to stranger-child dyads (IBD = 0.018, IBS = 0.375 ± 0.005, t (32) = 7.208, $p$ < .001, Fig. 3), when the children were correctly informed of their partner's identity. Furthermore, this maternal advantage could be modulated through partner-belief manipulation. The density of interbrain networks increased when children believed they were cooperating with their mother rather than a stranger, regardless of whether the actual partner was mother or stranger (IBD: *believed mother - actual mother* = 0.056, *believed stranger - actual mother* = 0.033, *believed mother - actual stranger* = 0.037, *believed stranger - actual stranger* = 0.018, Fig. 3B). The partner-belief effect on interbrain synchrony was further supported by a 2 × 2 repeated-measures ANOVA on IBS (*believed mother - actual mother* = 0.427 ± 0.005, *believed stranger - actual mother* = 0.406 ± 0.006, *believed mother - actual stranger* = 0.413 ± 0.004, *believed stranger - actual stranger* = 0.375 ± 0.005, Fig. 3C), which showed a significant main effect of *believed partner* (F (1,32) = 23.89, $p$ < .001, $\eta_p^2$ = .427) and a non-significant *believed partner* × *actual partner* interaction (F (1,32) = 3.36, $p$ = .076, $\eta_p^2$ = .095). However, the magnitude of partner-belief modulation of global IBS, quantified as the contrast between *believed mother* and *believed stranger* conditions under the same actual partner, did not show any significant correlation with secure attachment scores (both $|r|$ < .006, $p$ > .974). Beyond the believed partner effect, the main effect of actual partner was also significant (F (1,32) = 30.21, $p$ < .001, $\eta_p^2$ = .486), with mother-child dyads exhibiting higher interbrain synchrony than stranger-child dyads across both believed partner conditions.

It should be noted the maternal advantages — either the *actual partner* effect at side-by-side cooperation or the *believed partner* effect at remote setting — were only found in the theta band but not in the alpha band (Fig. S2 and Fig. S3). The results, consistent with our previous study[12], once again confirm the critical role of theta rhythm in child-actor/adult-observer interbrain communication.

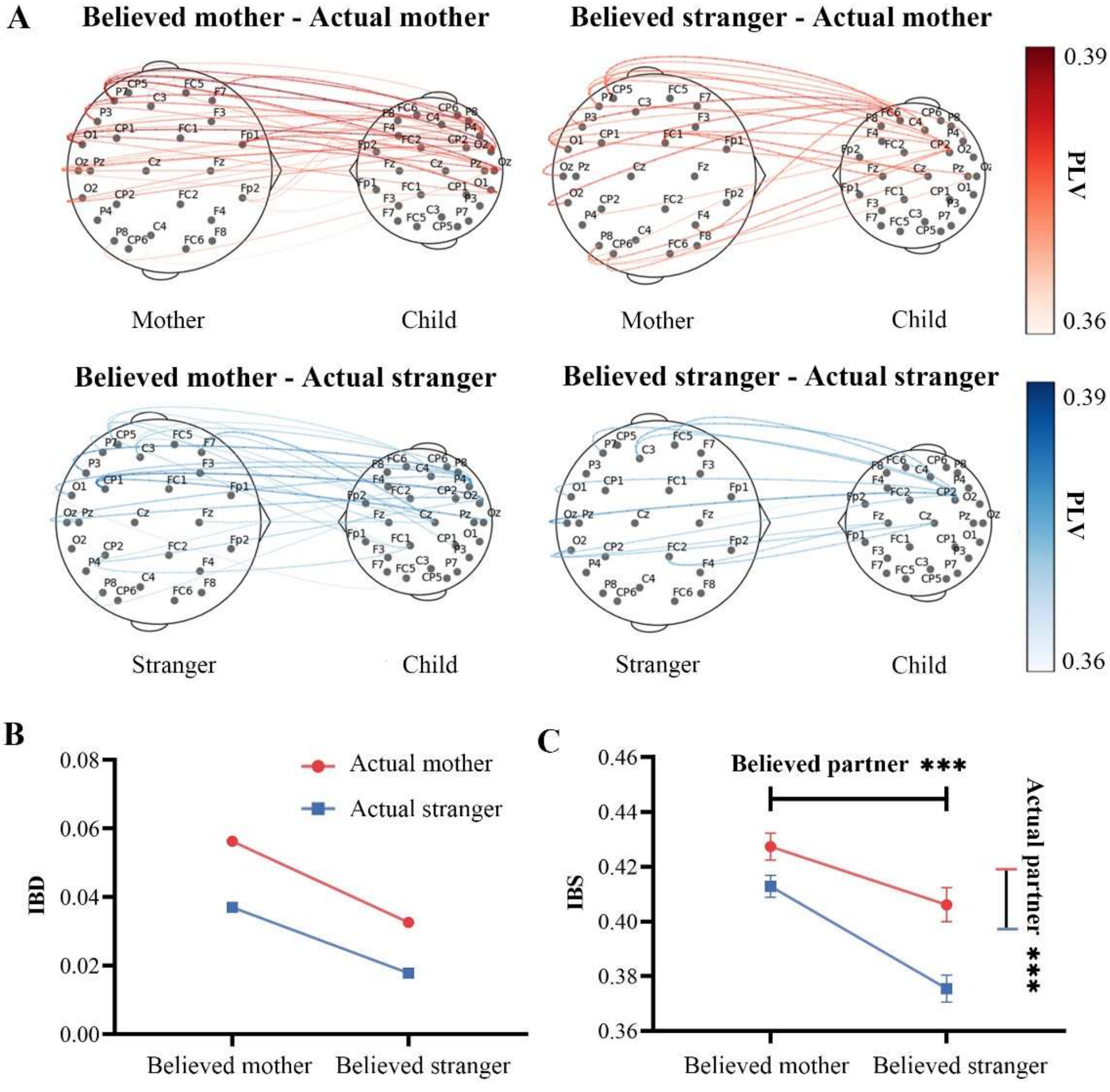

**Fig. 3 Global interbrain network metrics across different conditions.** (A) Interbrain network maps, (B) interbrain density (IBD), and (C) interbrain strength (IBS) in the theta band. $^{*}p$ <.05, $^{**}p$ <.01, $^{***}p$<.001.

**Local interbrain network characteristics**

To further explore the brain regions associated with child-to-mother attachment representations, we analyzed the changes in node degree of the theta-band interbrain networks induced by partner-belief manipulation (Fig. 4). Notably, the interbrain regions where the interbrain synchrony was sensitive to partner-belief manipulation were mainly found in child's right parieto-occipital region and adult's left centro-parieto-occipital region (Fig. 4A-B). In contrast to the sparse interbrain connectivity at P4, O2, and Oz electrodes during *believed stranger* conditions (*believed stranger - actual mother*: 1 connection, *believed stranger - actual stranger*: 0 connection), children exhibited numerous connections at these electrodes during *believed mother* conditions (*believed mother - actual mother*: 22 connections, *believed mother - actual stranger*: 11 connections). Of these, 6 connections (child_P4 - adult_C3, child_P4 - adult_CP1, child_P4 - adult_P3, child_P4 - adult_O1, child_P4 - adult_Oz, child_O2 - adult_Pz) emerged exclusively at the *believed mother* conditions but dissipated when *believed stranger*, manifesting as neural signatures of mother-partner belief (Fig. 4C). Critically, 5 of these connections were linked to child_P4 channel, and their mean strength at *believed mother - actual mother* condition was significantly correlated with both secure attachment scores ($\rho$ = .405, $p$ = .019, 95% CI: [.049, .644], Fig. 4D) and the mother-belief effect on reaction time ($\rho$ = .393, $p$ = .024, 95% CI: [.061, .651], Fig. 4E). That means, children who exhibited a greater reduction in children's reaction time from *believed stranger - actual mother* to *believed mother - actual mother* condition and higher secure attachment scores, demonstrated stronger IBS at P4 under *believed mother – actual mother* condition.

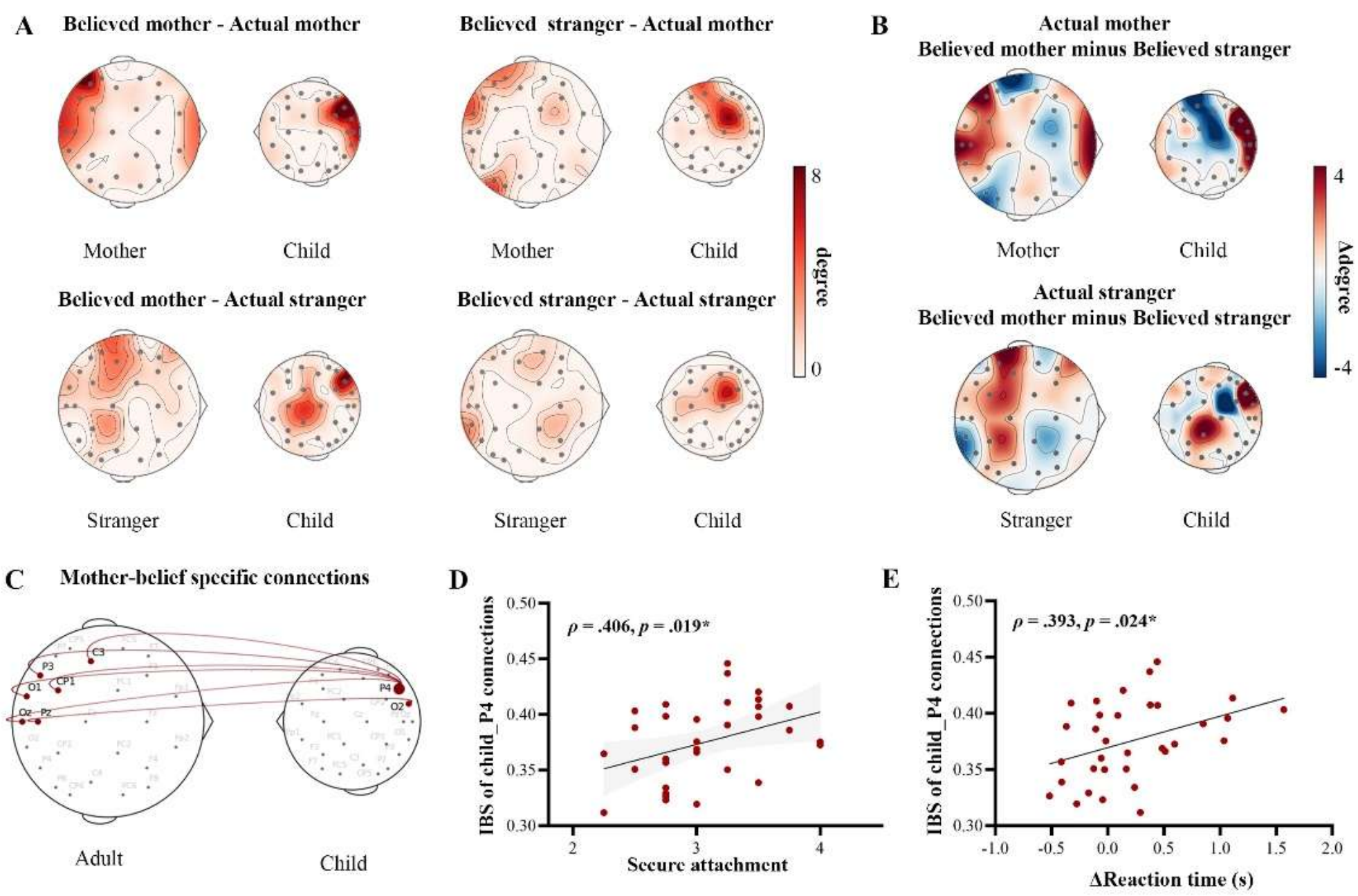

**Fig. 4 Modulation of local interbrain network features by partner belief.** (A) Topographies of node degree in different conditions. (B) The difference of node degree between *believed mother* and *believed stranger* conditions. (C) 6 mother-belief specific connections that emerge at the *believed mother* conditions but dissipate at *believed stranger* conditions. (D) The positive correlation between the mean IBS of child_P4 connections at *believed mother - actual mother* condition and secure attachment scores. (E) The significant positive correlation between Δ Reaction time (*believed stranger - actual mother* minus *believed mother - actual mother*) and the mean IBS of child_P4 connections at *believed mother – actual mother* condition. $^{*}p$ <.05.

# Discussion

The attachment representations not only sustain the human attachment system across the lifespan by serving as an emotional safe haven[5,32,33], but also extend the neurobiological template to diverse bonds, enabling uniquely human capacities for forming enduring relationships that transcend physical presence[4]. Yet, how these representations are instantiated in the brain and whether they independently shape real-time interactions remained unexplored. Through a delicately designed Remote Partner-Belief Manipulation paradigm, the current study successfully isolated internalized attachment representations from all sensory, behavioral, and identity confounders, and made them observable via interbrain synchrony.

First, to minimize the confounders from social cues inherent in the face-to-face interactions, we adopted a remote interactive setting, and observed that mother-child dyads still exhibited significantly greater interbrain synchrony than stranger-child dyads even when overt social cues were minimized. This finding supports the view that interbrain synchrony is not merely driven by the entrainment of overt social signals[10,14,15], but is also modulated by higher-order internal psychological processes, such as genetic relatedness[20,21], shared knowledge[22], common memories[23], as well as attachment representations[12,34]. These factors, characterized by their stability and long-term nature, remain effective even in the absence of physical proximity.

To further causally isolate the unique contribution of child-to-mother attachment representations from other partner-specific confounders, we employed an orthogonal design manipulating *believed partner* and *actual partner*. The significant main effect of *actual partner* demonstrated that genetic relatedness and endogenous factors accumulated through long-term mother–child interactions, such as shared experiences and common memories, do contribute to enhanced interbrain synchrony. More critically, regardless of whether the actual partner was the mother or a stranger, when the internal state of "being with mum" was symbolically activated, the child's brain shifted into a distinctive mode of neural coordination, selectively enhancing interbrain synchrony with the partner. Such real-time switching supports the theoretical hypothesis that internalized attachment representations are online regulatory schemas that proactively organizes neural alignment between interacting partners[35,36]. Unlike attachment processes in most mammals that rely predominantly on immediate sensory input[1], our findings highlight the enduring nature of human attachment representations beyond the physical absence of the attachment figure. Previous research in adults has demonstrated that attachment representations can be activated through symbolic cues (e.g. linguistic prompts), which was beneficial to the regulation of stress responses and emotional states[37,38]. Extending previous findings, our study demonstrated that even preschool-aged children are capable of engaging such abstract psychological representations to achieve neural coordination.

Regional interbrain network analysis provided further evidence that experimental manipulation-induced changes in interbrain synchrony were attachment-associated. Specifically, the P4 channel in children emerged as a critical node sensitive to partner-belief manipulation, with its IBS at *believed mother – actual mother* condition showing a positive correlation with secure attachment scores. Although EEG lacks the spatial resolution for precise anatomical localization, the P4 channel exhibits topographical proximity to right temporo-parietal junction region (rTPJ). Previous research has consistently highlighted the role of the TPJ during parent-child cooperative tasks, where its activation is interpreted as reflecting the representation of mental states and enhanced attentional allocation to social cues[39–42]. Critically, the rTPJ has been strongly implicated in processing emotional closeness and facilitating affiliative behaviors[43,44] — functionally dissociating it from the left TPJ that predominantly subserves linguistic computation and semantic integration[45,46]. This hemispheric asymmetry aligns with the right-lateralized dominance in mediating the social and biological functions of the attachment system[47,48]. Together, these findings suggest that attachment may tune neural activity within social-cognitive networks centered on the rTPJ toward a state that facilitates alignment and coordination with interactive partners.

One plausible mechanism underlying this attachment-facilitated interbrain synchrony is children's enhanced attention toward the attachment figure. In our study, children's attentional engagement was indexed by reaction time, a measure commonly employed to index one's focus toward an interaction partner[49,50]. Thus, the result that children exhibited accelerated reaction times when believing they were cooperating with their mother revealed a preferential attentional bias as well as enhanced task engagement toward the *believed mother* partner. This pattern is consistent with eye-tracking findings showing more gaze to mothers versus unfamiliar women[51], and align with attachment framework proposing that attachment-related priors for the mother bias the child to allocate greater perceptual, cognitive, and attentional resources to her[36,52]. We further linked the change in reaction time induced by partner-belief manipulation to the local IBS of children's P4 channel. Incorporating behavioral and neural findings, it can be inferred that attachment-related priors regulate children's online attentional engagement and manifest as higher interbrain synchrony, which may constitute an internal mechanism supporting more efficient communication, cooperation, and mutual understanding.

This study advances attachment theory by providing an operationally definable and empirically testable framework for the stable, enduring neurobiological signature of attachment representations. While previous research has consistently emphasized the crucial role of internal working model in bridging early attachment experiences and long-term developmental outcomes[4,33], its abstract and latent nature rendered it difficult to be directly assessed and operationally defined, leading to persistent criticism that it "can too easily explain anything"[53,54]. By establishing a neurophysiologically grounded measurement, our work provides a way to observe the real-time consequences of the internal working model, enabling future research to systematically investigate its content, development, and assessment. Furthermore, practically, the findings suggest that internalized attachment representations may serve as a psychological resource for supporting children's regulation and social engagement, implying that activating the child's attachment representation of the mother may be beneficial in contexts where separation is unavoidable.

Our study remains some limitations. Firstly, the relative homogeneity of children in our sample — characterized by typical development, secure attachment, and high socioeconomic status (SES) — may contribute to the large effect size ($\eta_p^2 = .427$) for the *believed partner* effects. Future studies would benefit from examining these effects across a more diverse range of developmental trajectories, attachment classifications (e.g., insecure or disorganized styles), and family backgrounds to evaluate the generalizability and boundary conditions of the observed mechanism. Furthermore, attachment security was assessed using mother reported Child–Parent Attachment Relationship Scale, which may introduce rater-related variance and limit measurement specificity for preschoolers' attachment representations. Future work should incorporate child-centered measures (e.g., story-stem narratives or observer-rated assessments).

# Methods

## Participants

We first estimated the required sample size based on prior effect sizes. According to the effect size ($\eta_p^2$ = .10) reported in our previous study of a similar tangram task in side-by-side condition[12], we conservatively estimated our sample size using a smaller effect size ($\eta_p^2$ = .05) to account for the reduced synchrony in remote setting. A minimum sample size of 28 (trios) was calculated with an $\alpha$ error probability of 0.05 and statistical power (1-$\beta$) of 0.8. Given an estimated 30% attrition rate due to task non-completion and poor data quality, particularly heightened in younger cohorts, we recruited 40 preschool children and their biological mothers from Shanghai Children's Medical Center. Additionally, 40 female adults were recruited as stranger partners to cooperate with the children. Therefore, a child, along with his/her mother and a randomly assigned stranger, constituted a trio.

All the participants were recruited according to the following criteria: (1) children's ages should be between 3-4 year (taking into account both children's ability to follow the instructions to complete the tasks and the developmental stage when attachment to the mother is already established yet brief separations can be tolerated); (2) all the participants should be right-handed according to the Edinburgh handedness inventory, with normal or corrected-to-normal vision, no known history of neurological or psychiatric disorders; (3) children should be with normal IQ measured by Wechsler scale, and no history of developmental delay reported by their mothers; (4) children should be classified as secure attachment according to the Child Attachment Scale[55]; (5) dyads followed the instruction to move puzzles using their right hand across the whole task; and (6) mothers should be the children's primary caregiver.

Four trios were excluded due to children's failure to complete the entire experimental procedure, and the final sample included 36 child-mother-stranger trios. The child sample comprised 18 boys and 18 girls (age = 3.65 ± 0.22 years, M ± SD). Mothers had a mean age of 34.39 ± 3.80 years, and strangers had a mean age of 27.64 ± 5.39 years. The children demonstrated relatively high cognitive ability in the Wechsler test and most of them were from families with high socioeconomic status (see Table. S1). All the participants signed a written informed consent before the experiment (parents on behalf of their children). This study was approved by the Institutional Review Board of Shanghai Children's Medical Center (SCMCIRB-K2023099-1).

## Experimental paradigm

To validate the hypothesis, we designed a Remote Partner-Belief Manipulation paradigm encompassing two experimental variables (Fig. 1A): (1) *believed partner* (*believed mother vs. believed stranger*) and (2) *actual partner* (*actual mother vs. actual stranger*). A child, along with his/her mother or a stranger, was seated in two adjacent, sound-proofed rooms and cooperated remotely to complete the seven-piece tangram puzzle-solving video tasks[12]. The pairs in collaboration were instructed to solve the puzzle piece by piece in turns. During the experiment, the child was either correctly informed about their partner or misled into believing that their partner was the deceptive one. Specifically, when a child was told that he/she was cooperating with his/her mother, the child was either actually cooperating with his/her mother (*believed mother - actual mother*) or a stranger (*believed mother - actual stranger*). Vice versa, when a child was told that he/she was cooperating with a stranger, the child was actually cooperating with his/her mother (*believed stranger – actual mother*) or a stranger (*believed stranger - actual stranger*). This 2 × 2 factorial design enabled isolation of biological relationship effects from belief-modulated social expectations.

In the experimental procedure (Fig. 1B), each child was instructed to complete the tangram puzzle task with different partners (both his/her mother and a stranger, with the order counterbalanced across dyads). The experiment included 2 blocks (1 block per believed partner). At the beginning of each block, to reinforce the partner-belief manipulation, the child was asked to practice 2 puzzle tangrams with the informed partner before the formal tasks, during which a live video of the partner was shown with the explicit instructions: "You will now complete the following tasks with your mother/a lady". Then, after the practice, the live video was removed, and the main task period (8 puzzle tangrams per block) began. The child was told to continue working with the same partner seen in the video, but in reality, the actual partner (mother or stranger) was randomly assigned for each tangram, with half (4 tangrams) cooperating with mother and the other half (4 tangrams) with stranger.

Additionally, our previous exploratory study indicated a dynamic EEG frequency-shift of interbrain synchrony with observer-actor role transitions during mother-child tangram puzzle-solving task[12,27]. Given the direct implications of these findings for the EEG frequency-band selection in the current study, we replicated the side-by-side cooperation experiment (a total of 4 tangrams) on the current cohort to validate the role-dependent neural coordination in EEG frequency.

**Setup and data acquisition**

The experiment setup consisted of a remote game system and three EEG recording systems. In the remote game system, a custom-designed puzzle game that contains 4 tangram templates for practice and 16 puzzle tangram templates for formal tasks, was displayed from a computer on two 23.8-inch touch screens (NEWTAP Inc., China). This setup enabled participants to cooperate using the touch screens in real time, both in the same room and remotely.

Triadic EEG signals were recorded simultaneously using three 32-channel mobile EEG systems (NeuSen.W32, Neuracle, China) at a sample rate of 1000 Hz, with FCz as the reference electrode and AFz as the ground. Game events, including touching the screen prior to moving each puzzle piece and lifting off the screen upon the completion of each piece, were automatically sent from the game program to three EEG systems to ensure inter-system synchronization. Meanwhile, videos during the tasks were also recorded along with EEG recordings. To minimize the motion artifacts — a known confounder in interbrain synchrony quantification[56,57], the participants were instructed to avoid any unnecessary movements except designated puzzle-moving activities, and their motions were monitored throughout tasks. However, many children failed to strictly adhere to this instruction and exhibited hyperactive behaviors during the partner's turn (the observing phase), which was further confirmed by quantified motion quantity using an open-source human pose estimation tool OpenPose[58] (Fig. S1). Considering the potential impact of excessive movements on interbrain synchrony, we only used the EEG data in child-actor/adult-observer role context (i.e., when adults observed children moving the puzzles) for analysis.

**Task performance assessment**

Movement duration (time elapsed from the initiation to the successful placement of a geometric piece) and reaction time (time interval between the completion of partner's piece placement and the initiation of the next piece movement) of each trial was calculated from the game events. The performance of task under different conditions was quantified using their median values across trials, with trials showing extreme reaction times (<500 ms or >5000 ms) excluded from the reaction time analysis.

**Mother-child relationship measures**

Child-to-mother attachment was assessed using the Child Attachment Relationship Scale[55], a maternal-report questionnaire that capture the child's attachment relationship with the mother based on typical child-mother interactions. The scale comprises 21 items, each rated on a 4-point Likert scale from 1 (strongly disagree) to 4 (strongly agree) (see Table. S2 for details). It measures three dimensions: secure attachment, avoidant/disorganized attachment, and ambivalent attachment. Items within each dimension were summed to yield a total score for that dimension. Children's attachment style was classified according to their highest dimension score. The scale has been used in Chinese samples and demonstrates good measurement validity, with the secure attachment subscale positively associated with interbrain synchrony during mother–child interaction[12]. Here, we only included children of secure attachment style in this study, and their secure attachment scores were used for analysis.

**EEG Processing**

The EEG data was preprocessed using EEGLAB toolbox in MATLAB (MATLAB R2019b, The MathWorks Inc., Natick, MA)[59], consisting of the following procedures: (1) filtering the raw EEG signal using 1-30 Hz band-pass and 50 Hz notch filters; (2) excluding the peripheral temporal channels with potential muscular artifacts (T7, T8, Tp9, Tp10); (3) applying independent component analysis (ICA) to remove eye-movement, blink, and muscular artifacts; (4) further filtering the EEG data into children's theta (3-6 Hz) and alpha (6-9 Hz) bands, respectively, which have been consistently substantiated to be responsible for interbrain communication between preschool children and adults[13,15,60].

After preprocessing, EEG recordings were divided into segments corresponding to puzzle piece movement events during gameplay. Due to the variance in the duration of moving a puzzle piece differed among participants and trials, 1-s epoch around the middle of each movement was selected. Epochs meeting any of the following criteria were rejected: (1) amplitude excursions exceeding ±100μV; (2) left-hand movement; (3) speech production; (4) intense head movement; or (5) observable inattention. Similar to our previous study[12], the epochs in in child-actor/adult-observer were segmented. Three trios were excluded due to poor EEG quality, and finally, an average of 11.75 ± 0.19 epochs remained (out of 14 total epochs) for each trio in each experimental condition.

**Interbrain synchrony calculation**

PLV[29], which has been commonly used for interbrain synchrony measurement, was adopted in this study. Briefly, PLVs were computed between all the channel-pair combinations of the child's EEG channels and the adult's EEG channels for each interactive condition within both theta and alpha bands.

With the aim of minimizing spurious connections potentially stemming from shared exogenous inputs (e.g., comparable environmental conditions or sensory signals)[13,15,61], we employed a thresholding procedure as mentioned in our prior work to isolate "real" interbrain links[12]. Specifically, we calculated a significance threshold based on the average PLV obtained from permuted data (generated by trial shuffling and random pairing trials in each dyad for 1000 times) across conditions. Only connections with original PLVs exceeding this threshold (one-tailed paired t-tests, $p < .05$) were retained for further analysis. Then, globally, interbrain density (IBD, referring to the proportion of the preserved "real" connections relative to all channel pairs) and interbrain strength (IBS, average PLV value of $N$ strongest connections among the preserved connections, where $N$ corresponds to the minimum number of significant connections shared across all experimental conditions) were calculated. Meanwhile, locally, node degree (the number of edges connecting to a certain node) was calculated to identify important brain regions.

**Statistical analysis**

Two-way repeated measure ANOVA was conducted on behavior performance and global IBS, taking *believed partner* and *actual partner* as within-group factors. In addition, Pearson's $r$ (or Spearman's $\rho$ if the variables are non-normally distributed) was applied to explore the relationships among behavior, interbrain strength and mother-child relationship measures. Statistical significance was accepted for $p < .05$. The normality of variables was tested by Shapiro - Wilk's test. Descriptive statistics for demographic variables are reported as mean ± standard deviation (SD) or n (%). All other experimental measures are presented as mean ± standard error of the mean (SEM). All the statistical analyses were conducted using Jamovi 2.3.21.

# Supplementary material

**Table. S1. Demographic characteristics of child-mother dyads**

| | Mean (SD)/N (Proportion) |
|---|---|
| **Characteristics of child–mother dyads** | |
| Child age (years) | 3.65 (0.22) |
| Child sex | |
| Boy | 18 (50.00%) |
| Girl | 18 (50.00%) |
| Child full scale IQ | 116.47 (9.12) |
| Maternal age (years) | 34.39 (3.80) |
| Maternal educational attainment | |
| Junior college or below | 2 (5.56%) |
| Undergraduate | 21 (58.33%) |
| Master or above | 13 (36.11%) |
| Annual household income (RMB) | |
| 150~300k | 10 (27.78%) |
| ≥300k | 22 (61.11%) |
| Unknown or refused to answer | 4 (11.11%) |
| **Characteristics of strangers** | |
| Stranger age (years) | 27.64 (5.39) |
| Stranger educational attainment | |
| Undergraduate | 10 (27.78%) |
| Master or above | 26 (72.22%) |

Note: Full scale IQ was measured by the Chinese version of Wechsler Preschool and Primary Scale of Intelligence-Fourth Edition (WPPSI-IV)

**Table. S2. The items of Child Attachment Relationship Scale**

| | Items | Strongly disagree | Disagree | Agree | Strongly agree |
|---|---|---|---|---|---|
| 1 | My child enjoys following me closely wherever I go. | | | | |
| 2 | At home, when I talk to strangers, my child prefers me to approach him/her. | | | | |
| 3 | When I come back shortly after going out, my child doesn't seem particularly happy to see me. | | | | |
| 4 | When I return from being outside, my child displays signs of annoyance or irritability towards my closeness. | | | | |
| 5 | When I'm getting ready to leave, my child constantly asks where I'm going. | | | | |
| 6 | My child clings to me and searches around when he/she can't see me. | | | | |
| 7 | Sometimes, my child feels like nobody loves him/her. | | | | |
| 8 | My child tends to exclude other children he/she doesn't like. | | | | |
| 9 | Even when I'm right beside him/her, my child is still afraid of strangers. | | | | |
| 10 | My child appears more passive when interacting with others in the class. | | | | |
| 11 | My child often feels unimportant and inferior to others. | | | | |
| 12 | When I take my child outside, he/she ignores me and explores or plays by himself/herself. | | | | |
| 13 | When my child encounters difficulties, he/she doesn't come to me for help. | | | | |
| 14 | When I try to intervene and assist my child when he/she struggles in a game, his/her reaction is as if I'm trying to take his/her toy or interfere with his/her own activities. | | | | |
| 15 | My child tends to get angry easily and reacts in an intense manner. | | | | |
| 16 | Sometimes, my child interacts better with strangers. | | | | |
| 17 | When I try to get close to my child, he/she avoids me and shows signs of fear. | | | | |
| 18 | My child often has difficulty concentrating. | | | | |
| 19 | My child lacks patience. | | | | |
| 20 | When my child plays a new game, he/she often appears unsure of what to do. | | | | |
| 21 | My child has many playmates, but he/she often says he/she doesn't have any close friends. | | | | |

**Motion extraction and quantitative analysis**

Given the potential effect of motion on interbrain synchrony[1,2], we employed OpenPose (https://github.com/CMU-Perceptual-Computing-Lab/openpose) to analyze participants' movements by extracting the body posture key points frame-by-frame (Fig. S1A)[3]. Right elbow kinematics were analyzed given their tracking validity and motion representativeness in the task. Specifically, after meticulously reviewing and adjusting the key points in each frame, frame-wise displacement (Euclidean distance between consecutive frames) of right elbow was normalized to individual shoulder-elbow distance and averaged per experimental condition. Then, the mean value of the motion across conditions were further calculated to quantify the motion quantity.

From the videos of 28 trios with reliably detectable right elbow points, we found that during the observing phase, adults remained near-quiescent states, showing significantly lower motion quantity than during the moving phase (data not conforming to the normal distribution, observing = 0.012 ± 0.001, moving = 0.034 ± 0.005, Wilcoxon W (27) = 4.00, $p < .001$, $r_{rb}$= -0.980, Fig. S1B). In contrast, children presented comparable motion levels across observing and moving phases (data normally distributed, observing = 0.057 ± 0.006, moving = 0.068 ± 0.005, Paired t (27) = -1.56, $p$ = .131, *Cohen's d* = -0.294, Fig. S1C).

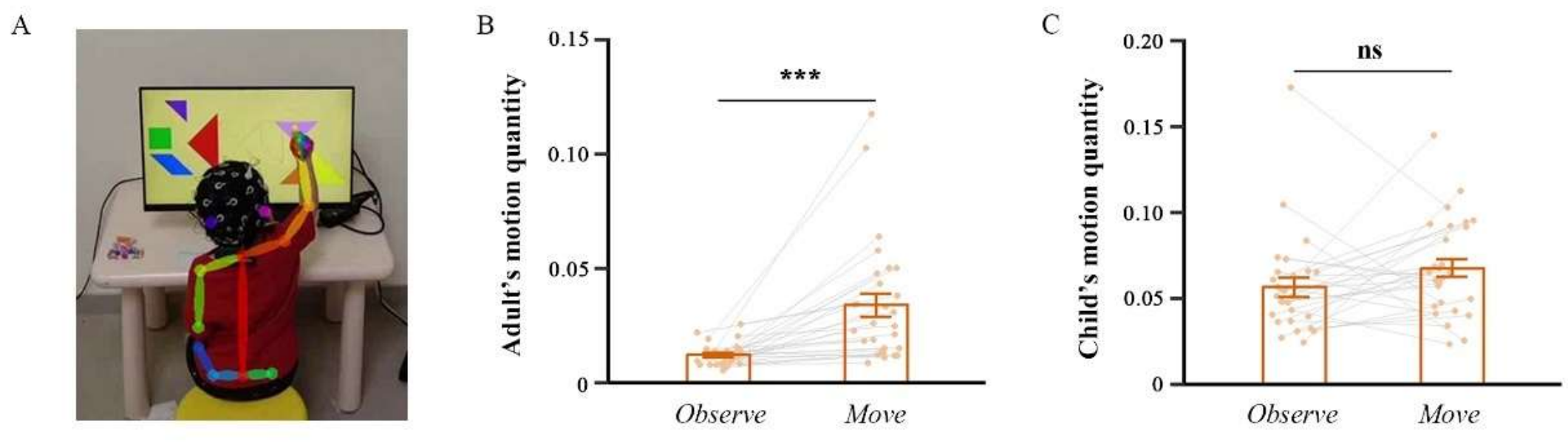

**Fig. S1. Motion assessment during tasks.** (A) An example of motion detection using OpenPose. (B) Mean value of motion quantity of children and adults when observing or moving the puzzles. Motion quantity was calculated as the average frame-wise displacement of right elbow (Euclidean distance between consecutive frames), normalized to individual shoulder-elbow distance. ***$p < .001$. ns. p>0.05.

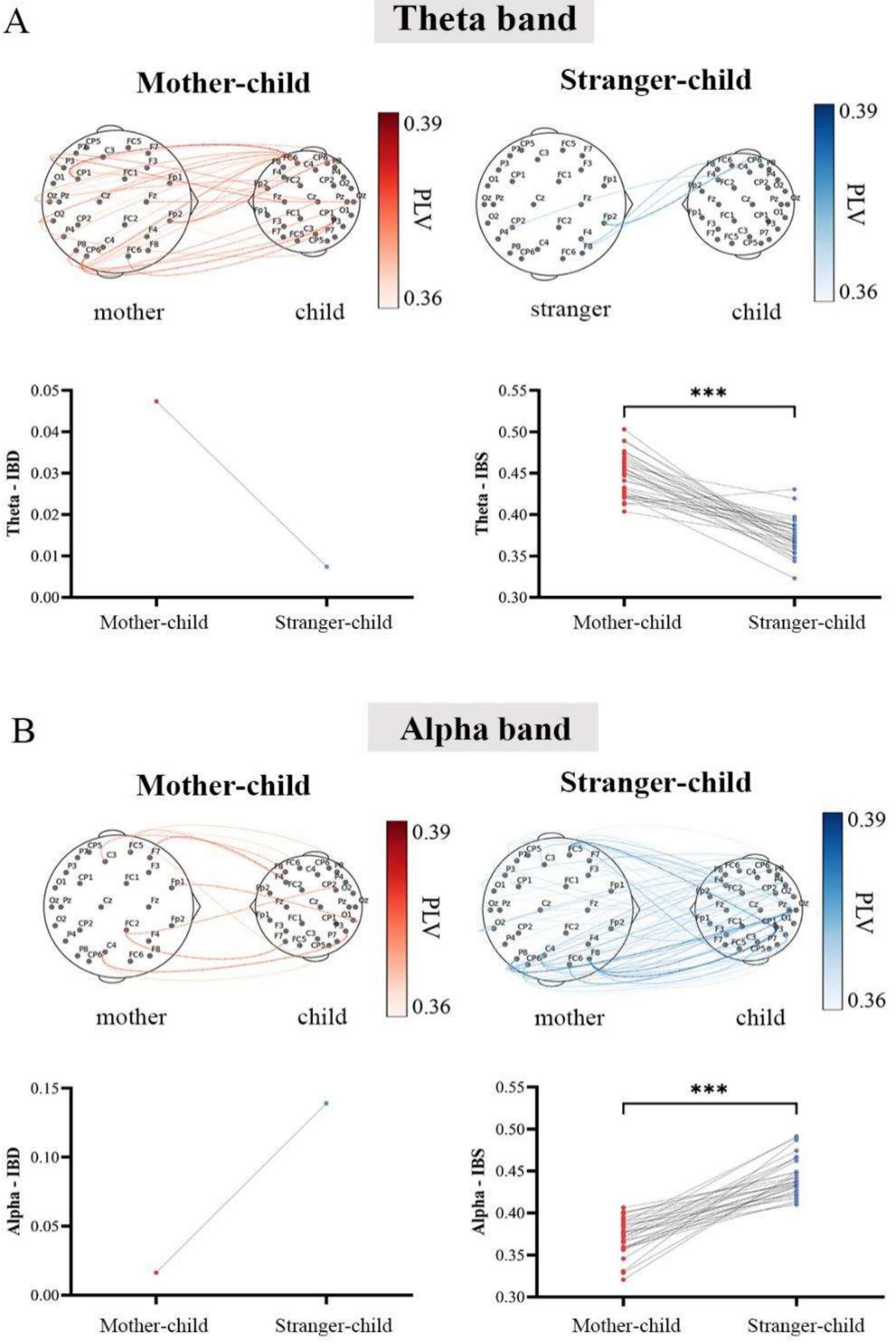

**Fig. S2. The neural synchrony difference when the child was cooperating with his/her mother and a stranger in the *side-by-side* condition.** When the adult partner was observing the child moving the puzzle, theta-band inter-brain synchrony was found to be denser and stronger in mother-child dyads(theta-IBD = 0.047, theta-IBS = 0.452±0.004) compared with stranger-child dyads (theta-IBD = 0.007, theta-IBS = 0.387±0.006, [t(32) = 9.72, $p$ <0.001]). However, in the alpha band, denser and stronger was found in stranger-child dyads (alpha-IBD = 0.016, alpha-IBS = 0.380±0.004) rather than mother-child dyads (alpha-IBD = 0.139, alpha-IBS = 0.439±0.004 [t(32) = -11.20, $p$ <0.001]). $^{*}p<.05$, $^{**}p<.01$, $^{***}p<.001$.

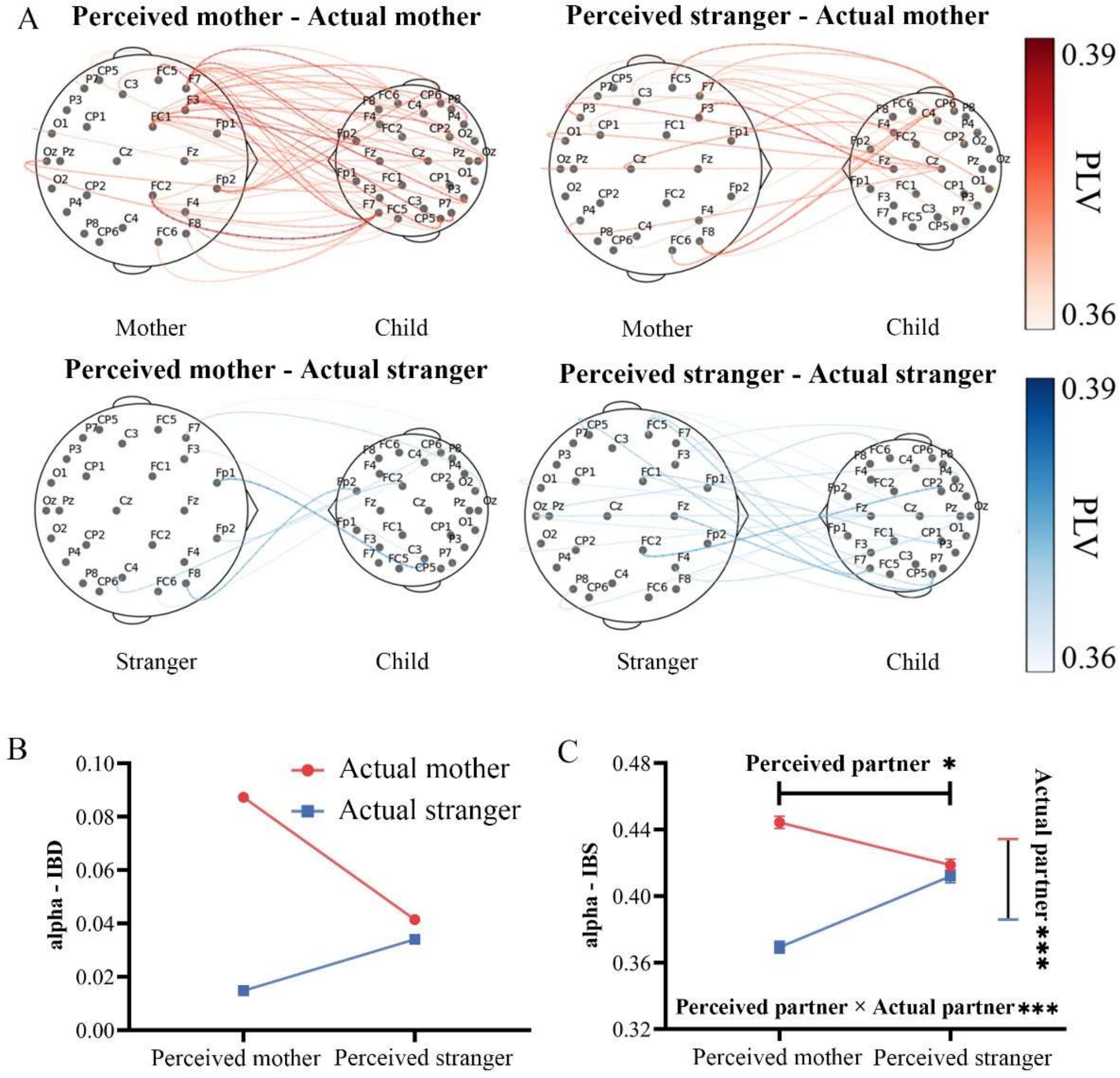

**Fig. S3. Modulation of global interbrain network metrics by partner belief in the alpha band.** (A) Interbrain network maps, (B) interbrain density (IBD), and (C) interbrain strength (IBS). We observed a significant interaction effect of perceived partner × actual partner (F (1,32) = 7.39, p = 0.01, $\eta_p^2$ =0.188) on global alpha-IBS. Although when actually cooperating with mother, there was a decrease from *perceived mother* to *perceived stranger* as expected (alpha-IBD: *perceived mother* = 0.087, *perceived stranger* = 0.041; alpha-IBS: *perceived mother* = 0.444±0.004, *perceived stranger* = 0.419±0.004 [t (32) = 5.82, p <0.001]), the increase from *perceived mother* to *perceived stranger* in actual stranger condition (alpha-IBD: *perceived mother* = 0.015, *perceived stranger* = 0.034,; alpha-IBS: *perceived mother* = 0.369±0.004, *perceived stranger* = 0.412±0.004 [t (32) = 5.82, p <0.001]) was contrary to our hypothesis. *p <.05, ***p<.001.